# Chip-scale atomic diffractive optical elements

*Liron Stern, Douglas Bopp, Susan Schima, Vincent N. Maurice and John Kitching*
*National Institute of Standards and Technology, Time & Frequency division, 325 Broadway Boulder CO, USA*

**Atomic systems have long provided a useful material platform with unique quantum properties. The efficient light-matter interaction in atomic vapors has led to numerous seminal scientific achievements including accurate and precise metrology[1–3] and quantum devices[4–7]. In the last few decades, the field of thin optical elements[8–12] with miniscule features has been extensively studied demonstrating an unprecedented ability to control photonic degrees of freedom, both linearly and non-linearly, with applications spanning from photography and spatial light modulators to cataract surgery implants. Hybridization of atoms with such thin devices may offer a new material system allowing traditional vapor cells with enhanced functionality. Here, we fabricate and demonstrate chip-scale, quantum diffractive optical elements which map atomic states to the spatial distribution of diffracted light. Two foundational diffractive elements, lamellar gratings and Fresnel lenses, are hybridized with atomic channels containing hot atomic vapors which demonstrate exceptionally strong frequency-dependent behaviors. Providing the design tools for chip-scale atomic diffractive optical elements develops a path for a variety of compact thin quantum-optical elements.**

Diffractive optical elements[12] and subwavelength dielectric-metallic elements (metasurfaces)[13] are important building blocks in science and technology. Such thin surfaces offer extraordinary control of the different degrees of freedom of light such as phase, polarization, and spectral and spatial distributions[14–16]. Moreover, devices can be tailored to be actuated mechanically[17], all-optically[18] and by polarization control[19]. Indeed, numerous optical elements have been implemented in this way including holograms[20], lenses[15,21] and Dammann gratings[22]. Atomic vapors are fundamental resources offering highly spectrally dependent and non-linear control of the phase and amplitude of light both in the quantum and classical regimes. Early combinations of atomic systems and thin surfaces have been demonstrated in the cases of dielectric[23] and metallic meta-surfaces[24] which effectively show that by combining such optical elements with standalone vapor cells or table-top vacuum chambers, hybridization of spatial modes, as well as polarization control can be achieved.

In this letter, we present a fully integrated chip-scale atomic diffractive optical element (ADOE). The ADOE consists of channels etched into silicon which are filled with gaseous alkali atoms forming an atomic-dielectric grating. The simplicity of the technique of anodic bonding glass over the silicon[5,6,25–27] lends itself well to wafer-level fabrication, providing dozens of devices in a single batch. By controlling the atomic state, the efficiency of the different diffraction orders can be tailored, thus mapping the atomic populations to the spatial diffraction pattern. The different spectra of the hybridized phase-amplitude atomic grating are measured and explained by a simple model. The hybrid element gains the steep dispersive properties in the vicinity of the atomic transitions while maintaining the rich designability of diffractive optical elements. Indeed, we further demonstrate an atomic Fresnel lens, capable of switching the efficiency of the focusing power by controlling the atomic state. Such a lens offers more than 95% contrast with potential atomic lifetime limited switching speeds in the mid-MHz regime. The fabrication process is highly scalable and may be easily adapted to incorporate metasurfaces or thin optical element with rubidium atoms incorporating all the atoms in the device behavior with minimal background contribution. As such, the concept may pave the way to a variety of quantum controlled, chip-scale thin elements.

In Figure 1a we present the concept of our ADOEs. Two types of diffractive optical devices are implemented. First, in figure 1a we illustrate a lamellar grating consisting of rectangular channels etched in silicon with rubidium vapor filling the channels and capped with borosilicate glass. The period of the diffraction grating is approximately 50µm and the etch depth is approximately 150µm. Also illustrated in Figure 1a is the diffraction pattern of light reflected from the grating surface. By scanning the frequency of the incident light, the diffraction efficiency is altered. For instance: two different detunings from the resonance frequency changes the overall phase response of the atomic system eliciting distinct spatial diffraction patterns illustrated in red and green (shifted for clarity). Figure 1b illustrates an atomic Fresnel lens. Here, a series of channels with varying period to facilitate lensing action is etched in silicon. An atomic reservoir is connected to these circular gratings via an additional etched channel. The focal plane is also illustrated in this figure, with two different states corresponding to two different frequencies: an on state, illustrated in red, and an off state illustrated in green.

A photograph of a typical device is presented in figure 1c where a few silicon based Fresnel lens are connected to an atomic reservoir which accommodates an alkali dispenser pill. The silicon frame (consisting of multiple optical elements) is anodically bonded to a borosilicate wafer and a $RbMbO_x$/AlZr pill is heat-activated using a laser to release natural abundance rubidium (see methods). Figure 1d presents a photograph of three typical devices after dicing.

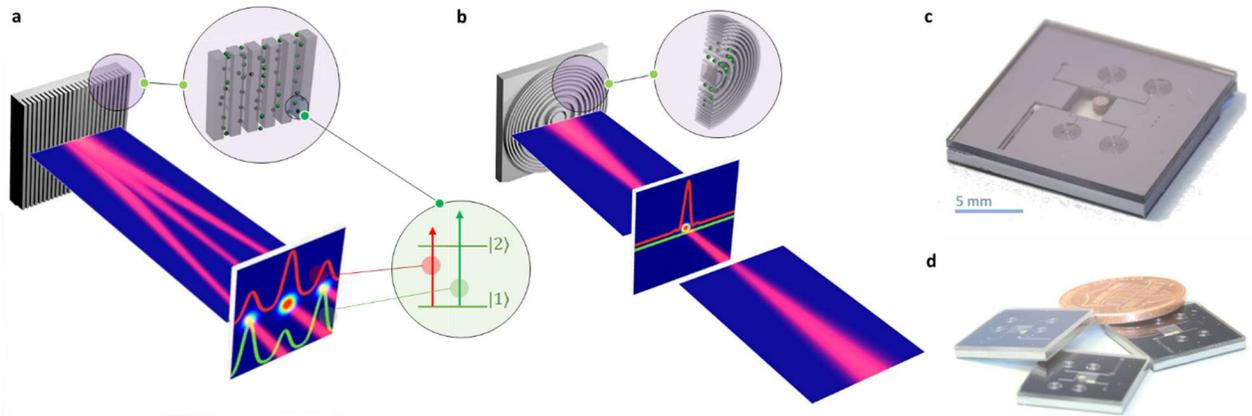

**Figure 1 | Atomic diffraction optical elements concept of operation** a) Artistic rendition of a diffractive atomic grating whose diffraction pattern is controlled by the atomic state of atoms embedded within its channels b) Artistic rendition of an atomic switchable Fresnel lens c) A photograph of a ADOE consisting of an rubidium reservoir connected to a few manifestations of Fresnel lens. d) A photograph of three typical diced devices compared to a penny.

We study the spectra of the atomic diffractive grating by first estimating the zeroth and first order modes as a function of frequency. By calculating[28] the rubidium bulk susceptibility and using it to introduce a phase-amplitude grating consisting of alternating columns of rubidium and silicon, we construct a theoretical diffraction response. Decomposing the bulk susceptibility into real and imaginary parts yields the amplitude and phase profiles, respectively, which are combined with the geometry of the device to fully define the phase-amplitude profile of the diffractive element. The frequency dependent phase-amplitude grating imposed by the rubidium atoms is modeled by Fourier transforming (i.e. $F(a(\omega,r) \cdot e^{i\phi(\omega,r)})$, where F is the Fourier operator, and $a$ and $\phi$ are the spatially and frequency dependent amplitude and phase response of the atoms, respectively) the phase-amplitude profile, from which we calculate the zero and first order diffraction spectra, plotted in Figure 2b. Figure 2a shows a reference absorption spectrum of the D2 line of natural rubidium. Marked with dashed lines are the peak positions of the Doppler broadened absorption line centers of $^{85}Rb$ and $^{87}Rb$; both spectra presented in Figure 2b exhibit strong oscillations in spectral regimes far from absorption peaks (i.e., between the two sets of doppler broadened absorption peaks, and red or blue detuned from either peak). On the contrary, in the regime of absorptive features, a flat saturated response is predicted due to the high optical density of the atoms preventing a frequency dependent response. The origin of such oscillations may be explained to be a result of the interferometric nature of the pure phase grating. In a symmetric square phase grating, the zero order diffraction intensity scales as the cosine of the phase difference. Moreover, as is also evident from Figure 2b, the first order diffraction spectrum is in quadrature to that of the zero order and exhibits an intensity that scales with the sine of the phase. Indeed, such calculations demonstrate the ability to control the efficiency of diffraction by means of controlling the phase response of the atomic medium, at a given frequency detuning.

Figure 2c plots the experimental first and zero order spectra, measured at a temperature of ~180⁰C and spatially resolved using a photodetector and a pinhole. The data was normalized to the maximum reflection of the zero-order as well as compensated for normally incident glass-air reflections. Generally, the fraction of reflected power to each of the orders was a few percent of the incident power. Clearly, the same features predicted by the simple two-level Fraunhofer model, presented in figure 2b, describe the experimental data. Interestingly, the experimental data exhibits small peaks and dips (which depended on the order of diffraction) within the presumably saturated absorption bands. The origin of such peaks is not fully understood and most likely stems from a contribution of atoms near the front borosilicate window forming an atom-dielectric response.

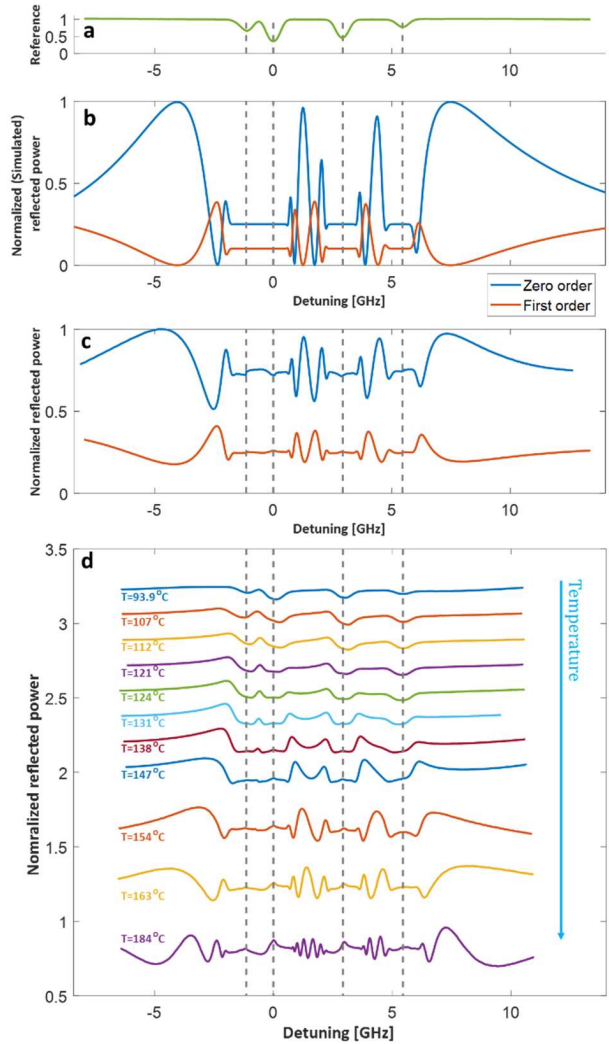

**Figure 2 | Atomic diffraction grating.** (a) Reference D2 rubidium absorption spectrum. The 0 GHz detuning is in reference to the $^{85}$Rb F=3 state. (b-c) Spectra of zero order (blue) and first order (orange) atomic diffraction grating b) calculated c) measured spectra d) evolution of the measured first order spectra as a function of atomic density.

The diffraction efficiency can be modulated by up to 50% by altering the laser frequency. This change is a pure phase change (with a calculated absorption of less than 0.5%) as it is occurring at a detuning of 5 GHz to the red of the $^{85}$Rb transition. The reduced contrast in the fringes is almost certainly due to the non-uniformity of etch depth (see methods) which gives rise to a loss of spatial coherence of the reflected light. Following, in figure 2d, we demonstrate the evolution of the first order diffraction spectrum as function of atomic density. The atomic density is varied from a density of ~$3\cdot10^{12}$cm$^{-3}$, (corresponding to a temperature of ~94$^0$C) to a density of ~$4\cdot10^{14}$cm$^{-3}$ (corresponding to a temperature of ~184$^0$C). We have observed that higher densities and high laser intensities are readily achievable and demonstrate exotic effects such as energy pooling and radiative collisions[29] which may be useful for indirect frequency conversion. With the lower density, the first order diffraction spectrum follows a typical absorption spectrum. By gradually increasing the atomic density, the spectra become more dispersive with spectral "wings" appearing between absorption bands finally evolving into oscillating patterns as discussed with respect to figures 2b and 2c. Interestingly, the frequency of such oscillations is directly related to the group index of the atomic medium in the so-called slow light regime between absorption peaks. Indeed, such a relation between group index and these features has been studied in the context of slow light enhanced interferometry[30].

The second element we demonstrate is the atomic Fresnel lens. As previously discussed, the lens constitutes annular concentric silicon rings with rubidium filling the space between the rings. The lens, designed to have a focal point of 70mm, has a dimeter of 2mm and the distance between rings ranges from 40μm to 120μm. A photograph of our device, is presented in Figure 3a. Here, a few Fresnel lenses which differ in size and topology are implemented in the same device. A section of the photograph is shown in the same figure where the rubidium dispenser connecting channels and two Fresnel lenses are visible.

We illuminate the lens with 780nm light resonant with the D2 resonances in rubidium and scan the laser about the absorption bands. We position a CCD camera at the measured focal plane of the Fresnel lens in absence of atoms. Figure 3a illustrates the optical power incident on a single pixel as a function of frequency detuning relative to $^{85}$Rb strongest resonance dip. The single pixel is chosen to coincide with the maximal power incident on the CCD. Figure 3b illustrates a series of cross-sections of the optical power profile incident on the CCD as the distance between the atomic lens and the CCD is changed. Figure 3c and 3d illustrate two different frequency detunings, -3.4GHz and --2.4GHz, respectively, which strongly modify the properties of the lens and effectively switch between focusing and not focusing the beam at the designed focal plane. The contrast of the optical power at these two frequencies exceeds 13dB and is equally illustrated in the large optical power swings seen in Figure 3a. As can be expected, the grating behavior at high optical densities when the laser is on the atomic resonances is strongly absorbed and does not contribute to lensing action. Conversely, far off resonance, the steep dispersion of the atoms strongly modulates the phase difference between the different Fresnel zones which allows the lens to be switched between focusing and defocusing actions.

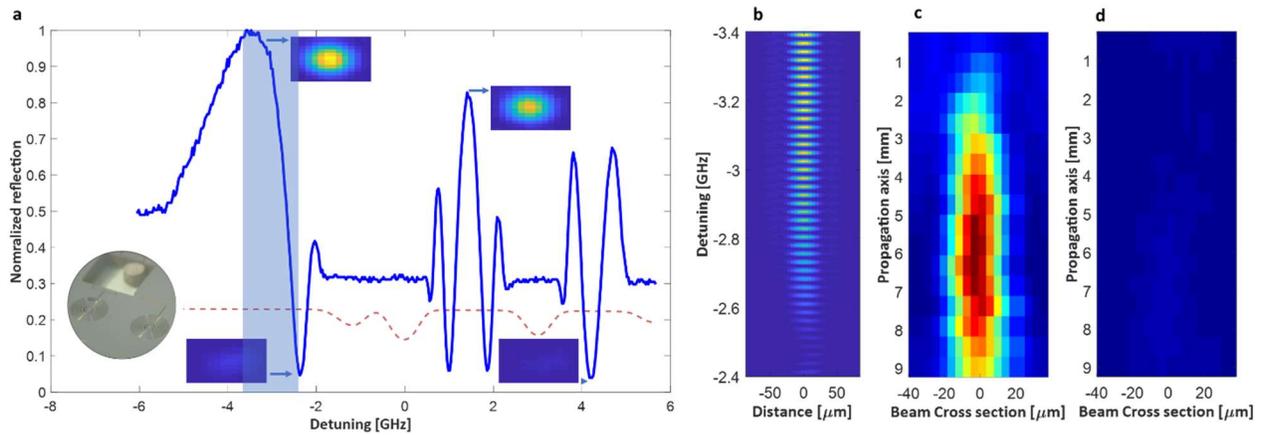

**Figure 3 | Atomic Fresnel lens** a) Fresnel lens spectrum at the focal point, obtained by recording a set of CCD images while simultaneously scanning the lasers frequency. Inset images are examples of such images at different frequency detunings. An additional inset shows a zoomed photograph of the actual device b) A series of images corresponding to the maximal position at -3.4GHz to the minimal point at -2.4GHz. c-d) The spatial dependence of the focal point cross section as function of propagation axis plotted for the two extreme c) "on" and d) "off" states.

To summarize, we have presented the concept of chip-scale, quantum diffractive optical elements. Consisting of traditional microscale diffraction optics embedded with rubidium vapor, the platform provides a unique system capable of mapping atomic properties onto the outputs of diffractive optical elements. We have demonstrated two distinct diffractive elements. The lamellar atomic grating demonstrates efficient control of the spatial distribution of light in the far field by controlling the phase of light interacting with the atoms within the grating. Moreover, we studied the atomic diffraction grating as functions of diffraction order, atomic density and frequency and found them to agree well with a simple model. By inheriting the same design features of the atomic diffraction grating, we demonstrate a quantum Fresnel lens. By characterizing the lenses' spatial and frequency dependent properties, we demonstrate efficient switching of the focusing properties of the lens with greater than 95% contrast by changing the frequency of the laser by only 1GHz

The incorporation of atomic vapors into DOEs offers unique properties to the overall ADOE, when compared to other tunable DOE material systems[31,32]. By encoding the atomic energy level structure into the photonic transfer function, one gains access to a fundamental material system offering atomically derived highly steep phase and amplitude gradients. Exploiting such properties already enables one to demonstrate intriguing applications spanning from highly-compact atomically derived frequency references and accurate optical spectrometers to advanced dispersive imaging systems. In this work we have exploited the frequency dependent refractive index change of the atomic medium yet other degrees of freedom such as chirality, entanglement, quantum coherence, non-linearity (allowing all optical control) and susceptibility to magnetic and electric fields atoms may further unravel new exciting applications and physics.

Providing the design tools for chip-scale atomic diffractive optical elements draws a path for a variety of compact, thin quantum-optical elements and quantum metasurfaces which may have an important impact on a myriad of fields within metrology and quantum technologies.

Note: Any mention of commercial products within this letter is for information only and does not represent an endorsement from NIST.

**Methods:**

**Device Fabrication and filling**
Two different processes were explored for wafer level device fabrication. The first, starting from a 2mm thick silicon 4" wafer, used deep reactive ion etching (DRIE) to blind etch the grating pattern into bulk silicon. Typical etch depths of 150μm were used in the devices reported here with aspect ratios of ~10. Following, a 1.5mm deep reservoir area was etched connecting to previously etched channels. This first process had a nonuniform etch depth due to the lack of an etch stop. With the second process, the same etching procedures were implemented exchanging the pure silicon wafer with a silicon on silicon-dioxide wafer. These wafers had three layers consisting of a 170μm thick silicon upper layer separated from an 830μm silicon handle by a 200nm silicon dioxide layer. Using the silicon-dioxide as an etch-stop layer, etch depth uniformity is significantly improved. Following etching, a wet buffered oxide etch removes the exposed silicon-dioxide. Finally, a second DRIE process is used to define the reservoir.

To introduce rubidium and seal the device, commercial rubidium dispenser pills consisting of a $RbMbO_x$/AlZr reducing agent[27] are added to the etched resevoirs.

Following, we bond a borosilicate wafer to the top (and bottom, for the case of the second type of process) of the silicon structure using a commercial anodic bonder (AML Wafer Bonder). We activate the dispenser pill to release rubidium into the cell by using approximately 1W of 980nm laser power weakly focused on the dispenser pill. Finally, the wafer is diced to achieve cm-scale lateral dimension devices.

**Optical Setup**

To characterize the ADOEs we use a 780nm DBR (Photodigm) laser operating at the D2 line of rubidium. Devices were heated using resistive heaters which generate small thermal gradients across the chip to avoid condensation of rubidium in the channel areas. Clogging of the channels was not observed in devices near the hot or cold side of the thermal gradient. A collimated laser beam of approximately 2mm diameter was directed to illuminate the specific device under test and the reflected diffracted beam is detected at normal incidence using a non-polarizing beam splitting cube directing light through a pinhole and onto a CCD camera or a photodiode. While scanning the laser across the D2 manifold, a reference cell spectrum is recorded.

**References:**

1. Budker, D. & Romalis, M. Optical magnetometry. *Nat. Phys.* **3,** 227–234 (2007).
2. Heavner, T. P. *et al.* First accuracy evaluation of NIST-F2. *Metrologia* **51,** 174–182 (2014).
3. Ramsey, N. F. A Molecular Beam Resonance Method with Separated Oscillating Fields. *Phys. Rev.* **78,** 695–699 (1950).
4. Kitching, J. Chip-Scale Atomic Devices. *Appl. Phys. Rev.* (2018).
5. Shah, V., Knappe, S., Schwindt, P. D. D. & Kitching, J. Subpicotesla atomic magnetometry with a microfabricated vapour cell. *Nat Phot.* **1,** 649–652 (2007).
6. Knappe, S. *et al.* A microfabricated atomic clock. *Appl. Phys. Lett.* **85,** 1460–1462 (2004).
7. Walker, T. G. & Larsen, M. S. Spin-Exchange-Pumped NMR Gyros. *Adv. At. Mol. Opt. Phys.* **65,** 373–401 (2016).
8. Zheludev, N. I. & Kivshar, Y. S. From metamaterials to metadevices. *Nat. Mater.* **11,** 917–924 (2012).
9. Segal, N., Keren-Zur, S., Hendler, N. & Ellenbogen, T. Controlling light with metamaterial-based nonlinear photonic crystals. *Nat. Photonics* **9,** 180–184 (2015).
10. Chen, H.-T., Taylor, A. J. & Yu, N. A review of metasurfaces: physics and applications. *Reports Prog. Phys.* **79,** 076401 (2016).
11. Brunner, R. Transferring diffractive optics from research to commercial applications: Part I – progress in the patent landscape. *Adv. Opt. Technol.* **2,** 351–359 (2013).
12. O'Shea, D. C., Suleski, T. J., Kathman, A. D., Prather, D. W. & Society of Photo-optical Instrumentation Engineers. *Diffractive optics : design, fabrication, and test.* (SPIE, 2004).
13. Yu, N. & Capasso, F. Flat optics with designer metasurfaces. *Nat. Mater.* **13,** 139–150 (2014).
14. Fattal, D., Li, J., Peng, Z., Fiorentino, M. & Beausoleil, R. G. Flat dielectric grating reflectors with focusing abilities. *Nat. Photonics* **4,** 466–470 (2010).
15. Lin, D., Fan, P., Hasman, E. & Brongersma, M. L. Dielectric gradient metasurface optical elements. *Science* **345,** 298–302 (2014).
16. Guo, C.-S., Yue, S.-J., Wang, X.-L., Ding, J. & Wang, H.-T. Polarization-selective diffractive optical elements with a twisted-nematic liquid-crystal display. *Appl. Opt.* **49,** 1069 (2010).
17. Li, X. *et al.* Stretchable Binary Fresnel Lens for Focus Tuning. *Sci. Rep.* **6,** 25348 (2016).
18. Papaioannou, M., Plum, E., Rogers, E. T. & Zheludev, N. I. All-optical dynamic focusing of light via coherent absorption in a plasmonic metasurface. *Light Sci. Appl.* **7,** 17157 (2018).
19. Gorodetski, Y., Niv, A., Kleiner, V. & Hasman, E. Observation of the Spin-Based Plasmonic Effect in Nanoscale Structures. *Phys. Rev. Lett.* **101,** 043903 (2008).
20. Macko, P. & Whelan, M. P. Fabrication of holographic diffractive optical elements for enhancing light collection from fluorescence-based biochips. *Opt. Lett.* **33,** 2614 (2008).
21. Deng, S. *et al.* Laser directed writing of flat lenses on buckypaper. *Nanoscale* **7,** 12405–12410 (2015).
22. Dammann, H. & Klotz, E. Coherent Optical Generation and Inspection of Two-dimensional Periodic Structures. *Opt. Acta Int. J. Opt.* **24,** 505–515 (1977).
23. Bar-David, J., Stern, L. & Levy, U. Dynamic Control over the Optical Transmission of Nanoscale Dielectric Metasurface by Alkali Vapors. *Nano Lett.* **17,** 1127–1131 (2017).
24. Aljunid, S. A. *et al.* Atomic Response in the Near-Field of Nanostructured Plasmonic Metamaterial. *Nano Lett.* **16,** 3137–3141 (2016).
25. Chutani, R. *et al.* Laser light routing in an elongated micromachined vapor cell with diffraction gratings for atomic clock applications. *Sci. Rep.* **5,** 14001 (2015).
26. Daschner, R. *et al.* Triple stack glass-to-glass anodic bonding for optogalvanic spectroscopy cells with electrical feedthroughs. *Appl. Phys. Lett.* **105,** 041107 (2014).
27. Hasegawa, M. *et al.* Microfabrication of cesium vapor cells with buffer gas for MEMS atomic clocks. *Sensors Actuators A Phys.* **167,** 594–601 (2011).
28. Siddons, P., Adams, C. S., Ge, C. & Hughes, I. G. Absolute absorption on rubidium D lines: comparison between theory and experiment. *J. Phys. B At. Mol. Opt. Phys.* **41,** 155004 (2008).
29. Kopystyńska, A. & Moi, L. Energy transfer in

collisions between excited atoms. *Phys. Rep.* **92,** 135–181 (1982).
30. Magaña-Loaiza, O. S. *et al.* Enhanced spectral sensitivity of a chip-scale photonic-crystal slow-light interferometer. *Opt. Lett.* **41,** 1431 (2016).
31. Komar, A. *et al.* Dynamic Beam Switching by Liquid Crystal Tunable Dielectric Metasurfaces. *ACS Photonics* **5,** 1742–1748 (2018).
32. Colburn, S., Zhan, A. & Majumdar, A. Tunable metasurfaces via subwavelength phase shifters with uniform amplitude. *Sci. Rep.* **7,** 40174 (2017).

**Acknowledgements**
The authors acknowledge Peter Lowell , Jim Nibarger, and James Beall for discussions and assistance in fabricating the devices, and Azure Hansen and James McGilligan for comments on the manuscript. This work is a contribution of the US government and is not subject to copyright in the United States of America.

# Chip-scale atomic diffractive optical elements – Supplementary

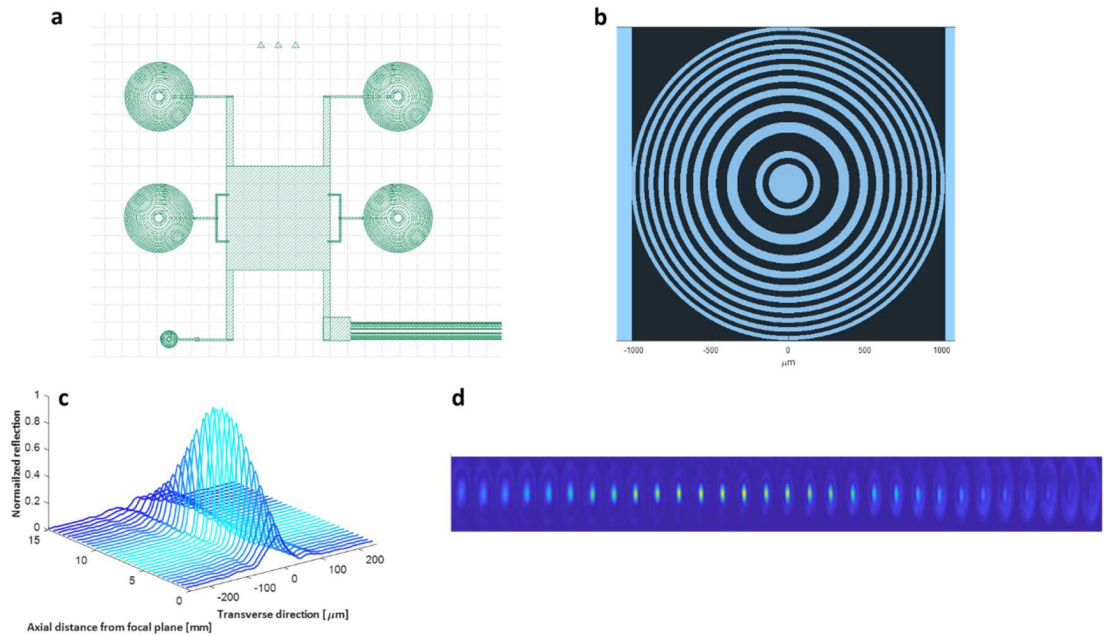

*Figure Supplementary 1: a) schematic layout of the device presented in figure 1c of the main manuscript b) typical dimensions and layout of the Fresnel lens b) normalized reflection profile as function of axial distance and one of the transverse directions of the Fresnel lens without active atoms d) Stacked images of the Fresnel lens across the focal point*

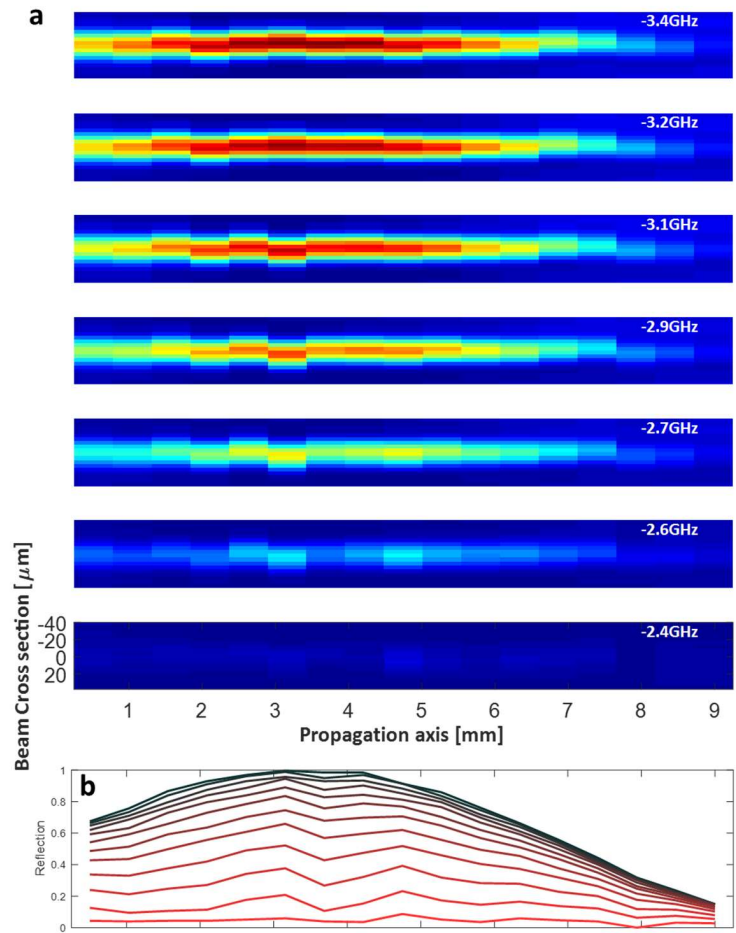

*Figure Supplementary 2:* a) Beam cross section as function of propagating axis, for different detuning (across 1GHz total detuning) b) The gradual transition of the lens operation around the center of axis for different frequency detuning's (corresponding to different phase responses of the atomic medium).

Supplementary 3 (Movie), this is a snapshot of the movie (attached as Media1.avi) :

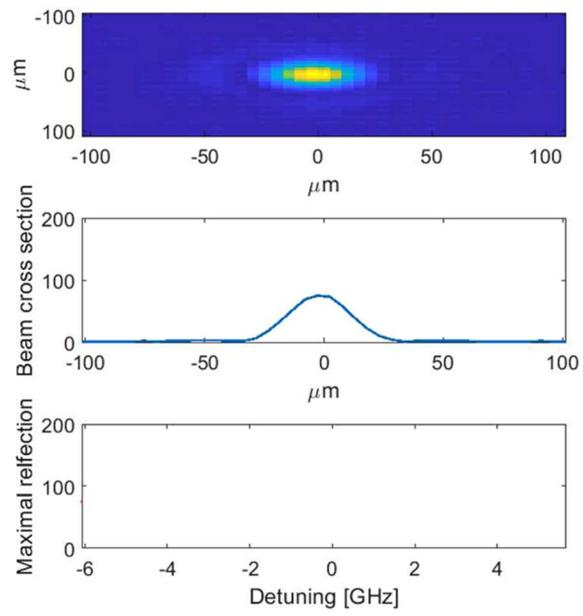